# On the limits of informationally efficient stock markets:
# New insights from a chartist-fundamentalist model

Laura Gardini[a,b], Davide Radi[c], Noemi Schmitt[d], Iryna Sushko[c], Frank Westerhoff[d*]

[a] Department of Economics, Society and Politics, University of Urbino Carlo Bo, Urbino, Italy
[b] Department of Finance, VSB-Technical University of Ostrava, Ostrava, Czech Republic
[c] DiMSEFA, Catholic University of the Sacred Heart, Milan, Italy
[d] Department of Economics, University of Bamberg, Bamberg, Germany

## Abstract

We utilize a chartist-fundamentalist model to examine the limits of informationally efficient stock markets. In our model, chartists are permanently active in the stock market, while fundamentalists trade only when their mispricing-dependent trading signals are strong. Our findings indicate the possible coexistence of two distinct regimes. Depending on the initial conditions, the stock market may exhibit either constant or oscillatory mispricing. Constant mispricing occurs when chartists remain the sole active speculators, causing the stock price to converge toward a nonfundamental value. Conversely, the stock price oscillates around its fundamental value when fundamentalists repeatedly enter and exit the market. Exogenous shocks result in intricate regime-switching dynamics.

___

[*] Contact: Frank Westerhoff, University of Bamberg, Department of Economics, Feldkirchenstrasse 21, 96045 Bamberg, Germany. Email: frank.westerhoff@uni-bamberg.de.

# 1 Introduction

The debate over whether stock markets are informationally efficient remains active among economists. Fama's (1970) "Efficient Market Hypothesis" asserts that stock markets are informationally efficient, meaning that stock prices reflect all available information. As a result, it is impossible to consistently achieve excess returns through technical or fundamental trading, as any new information that could impact prices is immediately and accurately incorporated into stock prices by rational market participants. In fact, Samuelson (1965) had already proved that informationally efficient stock prices are unforecastable and fluctuate randomly.

In contrast, Grossman and Stiglitz (1980) argue that if stock markets were perfectly efficient and stock prices always reflected all available information, there would be no incentive for speculators to gather and analyze information. Since information acquisition and analysis are costly activities, in a perfectly efficient market, the return on these activities would be negative because stock prices would already incorporate all information. This leads to a famous paradox: if no one gathers and analyses information because it is unprofitable, then stock prices cannot fully reflect all available information, thereby contradicting the Efficient Market Hypothesis. Consequently, stock markets must exhibit some form of mispricing to compensate speculators for the costs incurred in acquiring and processing information. Somewhat ironically, Grossman and Stiglitz's (1980) framework reveals an equilibrium degree of disequilibrium, characterized by constant mispricing.

According to Lo and Farmer (1999), Lo (2004), and Farmer (2024), the "Adaptive Market Hypothesis" offers a more comprehensive and realistic evolutionary perspective on the informational efficiency of stock markets. Since speculators are boundedly rational, heterogeneous, and influenced by cognitive biases, their collective actions can result in mispricing. Nonetheless, speculators are capable of learning and adapting their trading behavior over time. Consequently, stock market mispricing is not constant, and market efficiency should be viewed as dynamic rather than static.

Given that the determinants of market efficiency remain unclear, we propose a novel chartist-fundamentalist model to contribute to this important discussion. We identify conditions under which the stock price converges to a nonfundamental value, leading to constant mispricing. Under the same conditions, oscillatory mispricing may also occur, where the stock price fluctuates around its fundamental value. Coexisting attractors naturally accommodate both the insights of Grossman and Stiglitz (1980) and Lo and Farmer (1999) regarding the limits of informationally efficient markets within our model. We also find that stock prices may align with their fundamental values and evolve randomly, as envisioned by Fama (1970) and Samuelson (1965), although this occurs only under a very strict and unrealistic parameter configuration.

Recall that the chartist-fundamentalist approach, pioneered by Zeeman (1974), Beja and Goldman (1980), Day and Huang (1990), Lux (1995), and Brock and Hommes (1998), rests on the assumption that actual financial market participants rely on technical and fundamental analysis to predict stock prices. Menkhoff and Taylor (2007) and Hommes (2011) provide survey and experimental evidence supporting this assumption. As discussed by Murphy (1999), technical analysis seeks to derive trading signals from past stock price movements, a behavior that is often regarded as destabilizing. Fundamental analysis, reviewed by Graham and Dodd (1951), predicts that stock prices revert toward their fundamental values. Buying undervalued stocks and selling overvalued ones tends to stabilize stock market dynamics. Often, a market maker, who clears the market and adjusts the stock price in response to excess buying or selling, is included in these frameworks. This modeling device is also consistent with empirical evidence, see, e.g., Lillo et al. (2003) and Bouchaud et al. (2009). Chartist-fundamentalist models reveal that interactions between speculators using technical and fundamental trading rules may trigger complex boom-bust dynamics in stock markets. See Hommes (2013), Dieci and He (2018), and Axtell and Farmer (2024) for comprehensive literature reviews.

Our model adheres to the core principles of the chartist-fundamentalist approach, and follows Lo and Farmer's (1999) perspective on how financial markets function. Specifically, a market maker quotes the stock price based on current excess demand, chartists extrapolate past price trends into the future, and fundamentalists speculate on mean reversion. Our point of departure is that fundamentalists abstain from trading when their risk-adjusted profit expectations become negative. This assumption reflects the main argument presented by Grossman and Stiglitz (1980): speculators need an incentive to gather, analyze, and act on fundamental information. In our model, fundamentalists only follow their trading signals when they are strong enough to promise sufficient profit for their risk-taking. When their trading signals are too weak, they become inactive, leaving the stock market to the chartists' trading behavior. Fundamentalists' band of inactivity is centered around the stock market's fundamental value, which, in turn, follows a random walk. We are interested in how closely the stock price will track its fundamental value.

In the absence of exogenous shocks, the stock price in our model is driven by the iteration of a two-dimensional piecewise-linear discontinuous map. By employing a combination of analytical and numerical tools to explore this map, we establish the following results. The dynamics of our model depend on two parameters that measure the market impact of chartists and fundamentalists, allowing us to divide the corresponding parameter space into four distinct regions, labeled R1, R2, R3, and R4. Broadly speaking, we observe the following outcomes:

- In region R1, where the market impact of both chartists and fundamentalists is relatively low, the generic trajectory of the stock price always converges to a nonfundamental fixed point.
- In region R2, where the market impact of chartists and fundamentalists is moderate, the generic trajectory of the stock price either converges to a nonfundamental fixed point or exhibits endogenous dynamics.
- In region R3, where the market impact of fundamentalists is relatively high, the generic trajectory of the stock price either converges to a nonfundamental fixed point or displays divergent dynamics.
- In region R4, where the market impact of chartists is relatively high, the generic trajectory of the stock price is divergent.

The behavior of the stock market in regions R1 and R2 is of particular interest.

Consider region R1. Convergence to a nonfundamental fixed point implies constant mispricing. Since there is a continuum of fixed points within a band around the true fundamental value, the level of mispricing depends on the initial conditions. This convergence occurs because chartists are the only remaining speculators. As chartists do not receive trading signals prompting them to take a long or short position, the stock price remains unchanged. Although fundamentalists could theoretically trade on the mispricing, their expected risk-adjusted profits are negative. When the market impact of both chartists and fundamentalists is relatively low, this outcome represents the unique global behavior of our model.

In region R2, where the market impact of both is somewhat stronger but not too high, this regime may coexist with another regime that generates oscillatory stock market dynamics. Such dynamics arise from the interaction between chartists and fundamentalists, particularly from the repeated market entry and exit behavior of fundamentalists. For example, consider a scenario where the stock market is increasing but still quite undervalued. Then, both chartists and fundamentalists receive a buy signal, and their joint trading behavior results in a strong price increase. This, in turn, lends the stock market so much momentum that the chartists' trading behavior pushes the stock price considerably above its fundamental value. While fundamentalists may have temporarily exited the stock market, they now perceive profit opportunities and become active. As a result, the joint trading behavior of chartists and fundamentalists then leads to a significant downswing.

We are able to analytically characterize the basins of attraction of the stock market's nonfundamental fixed points and the magnitude of the stock price fluctuations. Policymakers may use our findings to design more efficient stock markets. The boundaries of these basins of attraction and the extent of the stock price fluctuations depend, among other factors, on the market impact of chartists and fundamentalists, as well as the risk-adjusted profit expectations held by fundamentalists. Policymakers may seek to influence these factors by implementing appropriate policies. To us, this seems to be highly relevant because such stock price dynamics occur for parameter settings that are usually associated with stable dynamics. Moreover, erratic switches between low-volatility periods with relatively constant mispricing and more volatile stock price dynamics – during which the stock price oscillates around its fundamental value – occur in region R2 in the presence of exogenous shocks. Knowledge of the basins of attraction of the coexisting attractors is crucial for understanding such dynamic behaviors.

A true highlight of our paper is the discovery of a new type of attractor, which we refer to as "weird quasiperiodic attractor". These attractors are neither periodic nor chaotic; they represent something in between. The attracting sets we observe in our model appear quasiperiodic: the trajectory comes as close as desired to each point but never revisits any previous point, i.e., they are not periodic. By showing that our map does not possess repelling cycles, we can also exclude that they are chaotic. However, the global shape of the attractor is difficult to predict and its form is weird, justifying our terminology.[1] Notably, this type of attractor is reminiscent to "strange nonchaotic attractors" found in systems with quasiperiodic forcing, as described by Feudel et al. (2006). However, strange nonchaotic attractors have fractal structures, motivating the term "strange". In contrast to "strange chaotic attractors", they do not display sensitivity to the initial conditions. Weird quasiperiodic attractors, in turn, emerge without any quasiperiodic forcing and it seems that they have no fractal structures.

The dynamics of our model are driven by a two-dimensional piecewise-linear discontinuous map. Such maps have recently been employed to study various phenomena in economics and finance, leading to several new insights. A key strength of these maps is their ability to provide clear-cut analytical insights into the functioning of economic systems, insights that are otherwise difficult to obtain. For studies using one-dimensional piecewise-linear maps, see Huang and Day (1993), Matsuyama (2007), Gardini et al. (2008), Tramontana et al. (2010, 2013), Matsuyama et al. (2016) and Jungeilges et al. (2021, 2022). Two-dimensional piecewise-linear maps are explored by Anufriev et al. (2020), Dieci et al. (2022), and Gardini et al. (2022, 2023, 2024a). For general surveys related to dynamic properties of nonsmooth maps and their applications in the social and natural sciences, we refer readers to Zhusubaliyev and Mosekilde (2003), Puu and Sushko (2006), di Bernardo et al. (2008), and Avrutin et al. (2019).

---

[1] In a more mathematically oriented companion paper, Gardini et al. (2024b) study the properties of weird quasiperiodic attractors in more detail.

We continue as follows. In Section 2, we develop our model and derive its law of motion. In Section 3, we explore the dynamics of our model. In Section 4, we conclude our paper. Appendices A and B contain proofs and further discussions.

## 2 Model setup and law of motion

Our model highlights the trading behavior of a market maker and two types of speculators: chartists and fundamentalists. The market maker determines excess demand, clears the market by taking an offsetting long or short position, and sets the stock price for the next period. Chartists believe in the persistence of price trends, demanding stocks when the price increases and supplying stocks when it decreases. Fundamentalists believe in mean reversion, demanding stocks when the market is undervalued and supplying stocks when it is overvalued. Importantly, fundamentalists trade only when their risk-adjusted profit expectations are positive, which is the case when their trading signals are sufficiently strong. Otherwise, fundamentalists become inactive. In such situations, the stock market is controlled by the trading behavior of chartists. The fundamental value of the stock market follows a random walk. Finally, there is also a nonspeculative demand for stocks that matches the total supply.

### 2.1 Model setup

Let us turn to the setup of our model. Similar to Hommes et al. (2005), Dieci et al. (2006), and Westerhoff (2012), we assume that a market maker adjusts the stock price based on current excess demand, using the linear price-adjustment rule

$$P_{t+1} = P_t + \alpha(D_t^C + D_t^F + D_t^R - N), \tag{1}$$

where $\alpha$ is a positive price adjustment parameter, and $D_t^C$ and $D_t^F$ represent the demands of chartists and fundamentalists. In addition, $D_t^R$ denotes the nonspeculative (real) demand for stocks, and $N$ is the total supply of stocks. For simplicity, we set the nonspeculative demand for stocks equal to the total supply of stocks, i.e.,

$$D_t^R = N, \tag{2}$$

resulting in a zero net supply of outside stocks. It follows that the market maker increases (decreases) the stock price when speculators' demand for stocks exceeds (fall short of) their supply of stocks.

To determine their speculative investment position in the stock market, chartists extrapolate past price trends into the future. See Murphy (1999) for a survey of technical trading rules. We formalize chartists' trading behavior as

$$D_t^C = \beta(P_t - P_{t-1}), \tag{3}$$

where $\beta$ is a positive reaction parameter reflecting chartists' market impact. Accordingly, chartists take a long (short) position in the stock market when the stock price increases (decreases). See Brock and Hommes (1998), Scholl et al. (2021), and Dieci et al. (2022) for related descriptions of the trading behavior of chartists.

Fundamentalists seek to exploit stock market mispricing. Mispricing in the stock market is determined by the distance between the stock price and its fundamental value $F_t$. Generally, fundamentalists enter a long position when the stock market is undervalued and a short position when it is overvalued. See Graham and Dodd (1951) for a classical treatment of fundamental analysis. The trading behavior of fundamentalists is often formalized as $D_t^F = \gamma(F_t - P_t)$, where $\gamma$ is a positive reaction parameter that captures the market impact of fundamentalists. See Brock and Hommes (1998), Scholl et al. (2021), and Gardini et al. (2022) for examples.

Grossman and Stiglitz (1980) argue that fundamentalists need an incentive to gather, analyze, and act on information. Given the complexity of determining an optimal market entry strategy, we assume that fundamentalists follow a simple rule of thumb, balancing the trade-off between expected profits and risk exposure. Specifically, fundamentalists abstain from the stock market when their expected risk-adjusted profits turn negative. Let fundamentalists' stock price expectations be given by $E_t^F[P_{t+1}] = P_t + \theta(F_t - P_t)$, with the expectation parameter $0 < \theta < 1$, so that they expect the stock price to change by $E_t^F[\Delta P_{t+1}] = \theta(F_t - P_t)$.

Fundamentalists consider two scenarios. First, when the stock market is undervalued, they expect the stock price to increase and are willing to enter a long position. Fundamentalists' risk-adjusted profit expectations are then $E_t^F[\pi_{t+1}] = E_t^F[\Delta P_{t+1}]D_t^F - \rho D_t^F$, where parameter $\rho > 0$ reflects their preference for risk compensation, which is proportional to their desired position in the stock market. Fundamentalists' expected risk-adjusted profits are positive if $F_t - P_t > \rho/\theta$. Second, when the stock market is overvalued, fundamentalists expect the stock price to decrease and are willing to enter a short position. Their risk-adjusted profit expectations, now given by $E_t^F[\pi_{t+1}] = E_t^F[\Delta P_{t+1}]D_t^F + \rho D_t^F$, are positive if $F_t - P_t < -\rho/\theta$. To simplify the notation, let us define $h \coloneqq \rho/\theta$. Parameter $h$ depends on fundamentalists' expectation formation behavior and risk attitudes.[2]

Combining these arguments, we formalize fundamentalists' trading behavior as:

$$D_t^F = \begin{cases} \gamma(F_t - P_t) & if \ P_t - F_t > h \\ 0 & if \ -h \leq P_t - F_t \leq h \\ \gamma(F_t - P_t) & if \ P_t - F_t < -h \end{cases}. \quad (4)$$

As before, parameter $\gamma > 0$ captures the market impact of fundamentalists, while parameter $h$ controls the critical level of the stock market's mispricing at which fundamentalists start to participate in stock market trading. As shown, fundamentalists only act on their trading signal when it is sufficiently strong, i.e., when they anticipate a profit that adequately compensates for their risk exposure. When their trading signal is too weak, fundamentalists become inactive.

We close our model by assuming that the fundamental value of the stock market follows a random walk, specified as

$$F_{t+1} = F_t + \delta_t, \quad (5)$$

where $\delta_t$ captures fundamental shocks, with $\delta_t \sim N(0, \sigma_\delta)$.

**2.2 Law of motion**

Combining (1) to (4) reveals that the evolution of the stock price of our model is

---

[2] Future work could explore more sophisticated market entry rules. For example, it would be interesting to examine the effects of asymmetric market entry levels. Moreover, parameter $\theta$ may also reflect trading costs that are proportional to fundamentalists' position in the market.

governed by

$$P_{t+1} = \begin{cases} (1 + \alpha\beta - \alpha\gamma)P_t - \alpha\beta P_{t-1} + \alpha\gamma F_t & \text{if } P_t - F_t > h \\ (1 + \alpha\beta)P_t - \alpha\beta P_{t-1} & \text{if } -h \leq P_t - F_t \leq h \\ (1 + \alpha\beta - \alpha\gamma)P_t - \alpha\beta P_{t-1} + \alpha\gamma F_t & \text{if } P_t - F_t < -h \end{cases} \quad (6)$$

Ideally, the stock price closely tracks its fundamental value. In an informationally efficient market, the trading behavior of the market participants would always ensure that $P_t = F_t$. Since we are interested in the relationship between the stock price and its fundamental value, we express our model in deviations from the fundamental value by defining $x_t = P_t - F_t$. For ease of exposition, we rename the aggregate parameters $\alpha\beta$ and $\alpha\gamma$ as $b$ and $c$, respectively. From (5) and (6), we then obtain

$$x_{t+1} = \begin{cases} (1 + b - c)x_t - bx_{t-1} + d_t & \text{if } x_t > h \\ (1 + b)x_t - bx_{t-1} + d_t & \text{if } -h \leq x_t \leq h, \\ (1 + b - c)x_t - bx_{t-1} + d_t & \text{if } x_t < -h \end{cases} \quad (7)$$

where $d_t = -\delta_t - b\delta_{t-1}$, with $d_t \sim N(0, \sigma_d)$ and $\sigma_d = \sigma_\delta \sqrt{1 + b^2}$.

Abstracting from exogenous shocks and introducing the auxiliary variable $y_t = x_{t-1}$ allows us to express the law of motion of the deterministic skeleton of our model in terms of the two-dimensional piecewise-linear discontinuous map

$$M: \begin{cases} x' = \begin{cases} f_R(x) = (1 + b - c)x - by & \text{if } x > h \\ f_M(x) = (1 + b)x - by & \text{if } -h \leq x \leq h \\ f_L(x) = (1 + b - c)x - by & \text{if } x < -h \end{cases} \\ y' = x \end{cases}, \quad (8)$$

where the prime symbol stands for the unit time advancement operator.

Map $M$ has the following properties. Recall that parameters $b$, $c$, and $h$ are positive. However, parameter $h$ is a scaling factor. By changing variables to $x \coloneqq x/h$ and $y \coloneqq y/h$, we can set $h = 1$. In this sense, the dynamics of our model depends only on parameters $b$ and $c$ and on the initial conditions $x_0$ and $y_0$. Due to its economic relevance, we retain parameter $h$ in our computations. Since $M(-x, -y) = -M(x, y)$, map $M$ is symmetric with respect to the origin. Therefore, any invariant set of map $M$ is either symmetric with respect to the origin, or there exists a corresponding set symmetric with respect to the origin. See Appendix B for a brief discussion.

## 3 Analytical and numerical results

### 3.1 Preliminary observations

Before we continue with the analysis of map $M$, let us define two special cases. For $h = \infty$, map $M$ becomes map

$$C: \begin{cases} x' = (1+b)x - by \\ y' = x \end{cases}. \tag{9}$$

Map $C$ describes the deterministic dynamics of the stock price in deviation from its fundamental value when chartists are the sole type of active speculator. For $h = 0$, map $M$ reduces to map

$$F: \begin{cases} x' = (1+b-c)x - by \\ y' = x \end{cases}. \tag{10}$$

Map $F$ captures the deterministic dynamics of the stock price in deviation from its fundamental value when chartists and fundamentalists are always jointly active.

What can we conclude about maps $C$ and $F$? Let us start with map $F$. Map $F$ has a unique fixed point, equal to the origin. At this fixed point, the stock market is informationally efficient, meaning that the stock price equals its fundamental value. The Jacobian matrix of map $F$ reads

$$J_F = \begin{bmatrix} 1+b-c & -b \\ 1 & 0 \end{bmatrix}, \tag{11}$$

with trace $tr = 1 + b - c$ and determinant $det = b$. The eigenvalues of the matrix are $\lambda_{1,2} = (1 + b - c \pm \sqrt{(1+b-c)^2 - 4b})/2$. These eigenvalues lie inside the unit circle, implying that the fixed point of map $F$ is globally stable, provided the following stability conditions are jointly satisfied: (i) $1 + tr + det > 0$, (ii) $1 - tr + det > 0$, and (iii) $1 - det > 0$. As can be easily verified, the stability domain of the fixed point of map $F$ is given by the stability box S, defined as $0 < c < 2 + 2b$ and $0 < b < 1$. The stability box S is depicted in green in the left panel of Figure 1. Within this stability box, the eigenvalues fall into three sub-regions: S1, S2, and S3, where the eigenvalues are complex conjugate, real and negative, and real and positive, respectively. Outside the stability box, in the gray region, the dynamics are divergent.

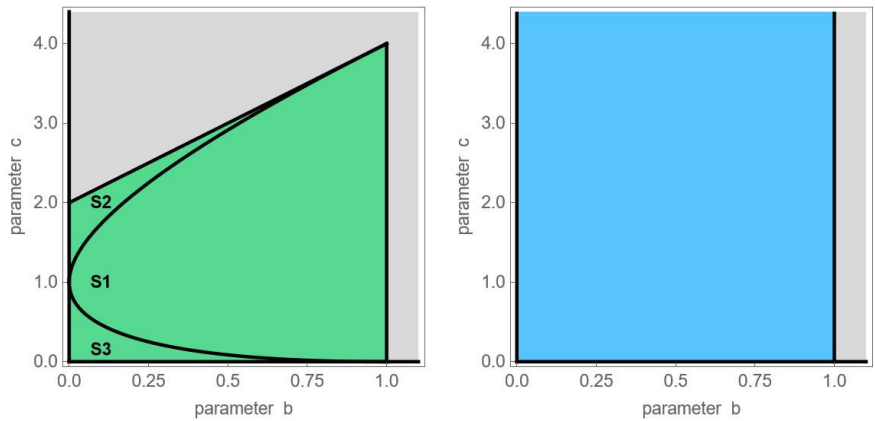

Figure 1: Stability domains of the fixed points of maps $F$ and $C$. The panels show the stability domains of the unique fixed point of map $F$ (left, green) and the continuum of fixed points of map $C$ (right, blue). Parameter combinations marked in gray result in divergent dynamics.

Figure 2 presents three examples of the dynamics of map $F$ in the time domain (left) and state space (right). The pink lines depict the stock price and the green lines their fundamental values. We will clarify the meaning of the black and blue lines in the sequel. The parameter setting in the top panels is $b = 0.80$ and $c = 1.35$. Since this parameter combination is located in sub-region S1, the stock price displays dampened oscillations around its fundamental value. The parameter setting in the middle panels, $b = 0.20$ and $c = 2.30$, is located in sub-region S2, resulting in an alternating adjustment path. The parameter setting in the bottom panels, $b = 0.20$ and $c = 0.30$, is located in sub-region S3, yielding a monotonic adjustment path.

To summarize the main implications of map $F$, as long as the reaction parameters of chartists and fundamentalists remain within the stability box S, their joint trading behavior eventually leads to informationally efficient stock markets. In other words, mispricing is only a temporary phenomenon. According to Fama (1970), stock prices always reflect their fundamental value. For the parameter constellation $b = 0$ and $c = 1$, we indeed observe that the stock price may consistently equal $P_t = F_t$. Since the fundamental value follows a random walk, so does the stock price – corresponding with Samuelson's (1965) prediction that informationally efficient stock prices are unforecastable and fluctuate randomly. However, this outcome depends on a strict and, given that it ignores the market impact of chartists, unrealistic parameter assumption.

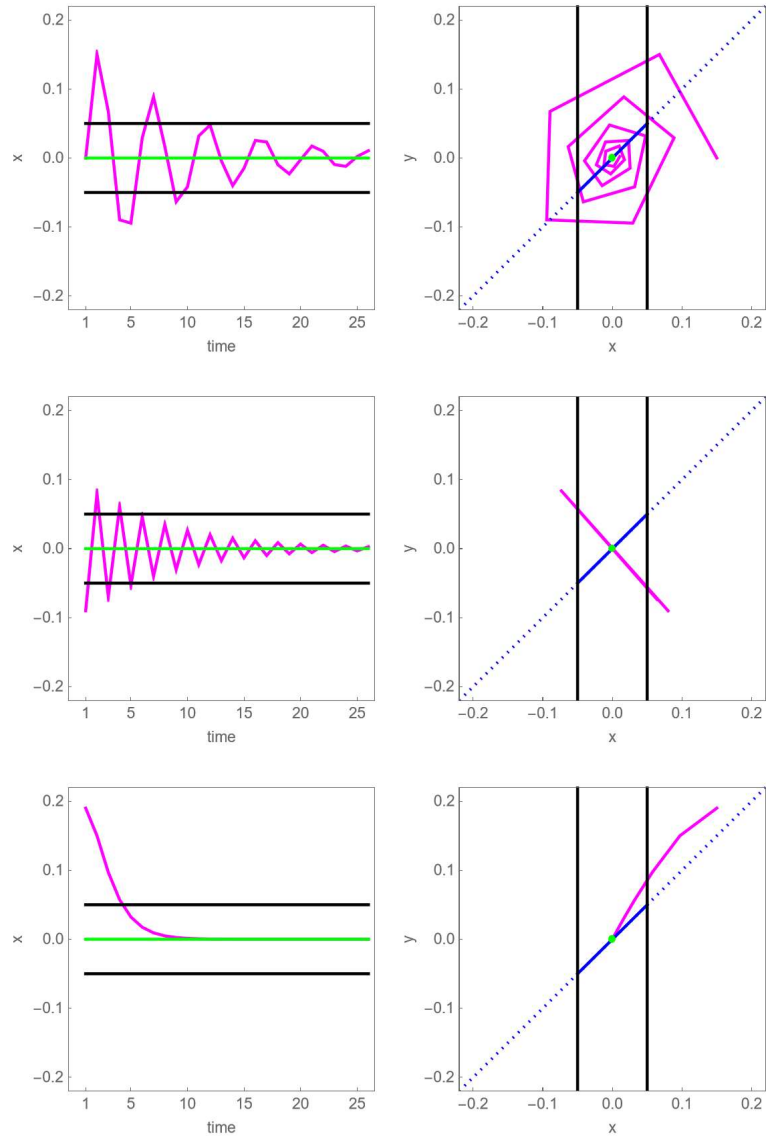

Figure 2: Dynamics of map $F$ in the time domain and in state space: Top: $b = 0.80$, $c = 1.35$, $x_0 = 0.15$, and $y_0 = 0.00$. Middle: $b = 0.20$, $c = 2.30$, $x_0 = 0.08$, and $y_0 = -0.09$. Bottom: $b = 0.20$, $c = 0.30$, $x_0 = 0.15$, and $y_0 = 0.19$.

Map $C$ has a continuum of fixed points, corresponding to the entire 45-degree line. At these fixed points, the stock market is mispriced, except at the origin. The Jacobian matrix of map $C$ is

$$J_C = \begin{bmatrix} 1 + b & -b \\ 1 & 0 \end{bmatrix}, \qquad (12)$$

with trace $tr = 1 + b$ and determinant $det = b$. Its two eigenvalues are $\lambda_1 = 1$ and $\lambda_2 = b$. The general solution of map $C$ is given by

$$G = \begin{cases} x_t = \frac{by_0-x_0}{b-1} - \frac{b(y_0-x_0)}{b-1}b^t \\ y_t = \frac{by_0-x_0}{b-1} - \frac{b(y_0-x_0)}{b-1}b^{t-1} \end{cases}. \tag{13}$$

Starting from the initial conditions $(x_0, y_0)$, the system ultimately converges to the fixed point $(\frac{by_0-x_0}{b-1}, \frac{by_0-x_0}{b-1})$ as long as $0 < b < 1$. In state space, the dynamics of map $C$ evolves along the line $y = \frac{x}{b} + \frac{by_0-x_0}{b}$. The blue region in the right panel of Figure 1 shows the stability domain of the continuum of fixed points of map $C$. For $b > 1$, the dynamics is divergent. This parameter region is depicted in gray.

Figure 3, based on $b = 0.80$, presents two examples of the dynamics of map $C$ in the time domain (left) and state space (right). Once again, the pink and green lines reflect the stock price and its fundamental value, respectively, while the meaning of the black and blue lines are clarified in the sequel. For the initial conditions $(-0.13, -0.17)$, the stock price moves towards the nonfundamental fixed point $(0.03, 0.03)$. For the initial conditions $(-0.10, -0.17)$, the stock price reaches the nonfundamental fixed point $(0.18, 0.18)$. Apparently, the stock market's final mispricing hinges on the initial conditions.

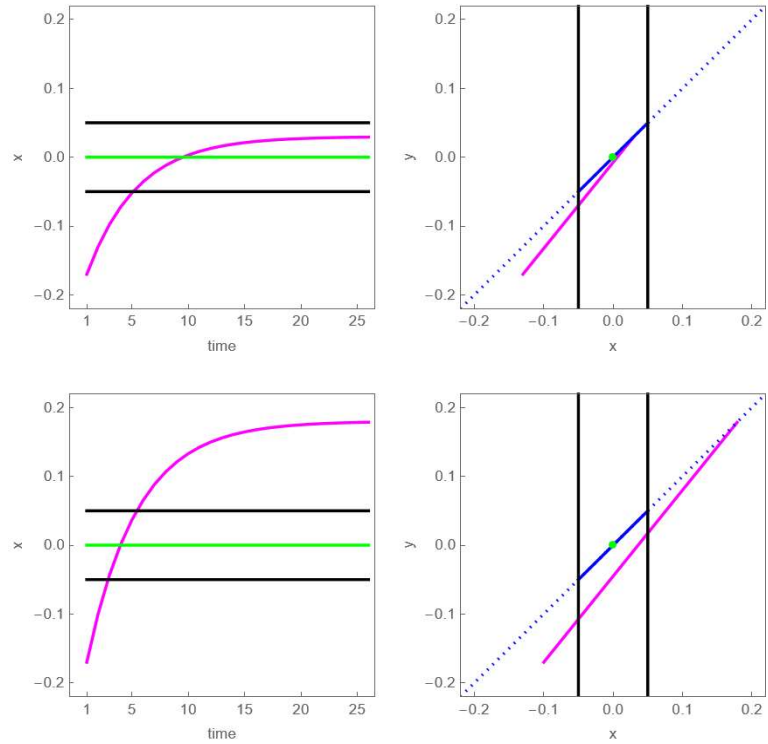

Figure 3: Dynamics of map $C$ in the time domain and in state space. Top: $b = 0.80$, $x_0 = -0.13$, and $y_0 = -0.17$. Bottom: $b = 0.80$, $x_0 = -0.10$, and $y_0 = -0.17$.

To summarize the main implications of map $C$, as long as $0 < b < 1$, all the initial conditions yield constant mispricing, except for those located on the line $y = x/b$. However, the final mispricing may be quite substantial, and in such cases, we would expect fundamentalists to seek to exploit it.

Our focus is not on the isolated dynamics of maps $F$ and $C$, but rather on their joint dynamics, represented by map $M$. In Figures 2 and 3, the black vertical and horizontal lines reflect the meaning of parameter $h$, set here to $h = 0.05$. According to our model, fundamentalists are inactive between the two lines defined by $x = h$ and $x = -h$. As a result, certain dynamics observed in Figures 2 and 3 are not compatible with map $M$. The dynamics of map $F$ are valid outside the black lines, while the dynamics of map $C$ are valid inside them. In the examples depicted in Figures 2 and 3, these boundaries have been violated. With these insights in hand, we are now prepared to examine map $M$.

Map $M$ has three branches, namely $f_L(x)$, $f_M(x)$, and $f_R(x)$. Their areas of definition are shown in the left panel of Figure 4 for $h = 0.05$. Due to its outer branches $f_R(x)$ and $f_L(x)$, map $M$ has two fixed points with identical coordinates at the origin. However, these fixed points are virtual since they exist outside the areas where their respective branches are defined. The Jacobian matrices of the branches $f_R(x)$ and $f_L(x)$ of map $M$ are identical to that of map $F$, i.e., $J_F = J_R = J_L$. In addition, due to its inner branch $f_M(x)$, map $M$ has a continuum of real fixed points with $-h \leq x = y \leq h$, represented by segment $S^*$. Only one of these fixed points, the origin, ensures informationally efficient stock markets. All other fixed points on segment $S^*$ result in constant mispricing. In the left panel of Figure 4, the segment $S^*$ of fixed points is marked in blue, with the origin highlighted in green. The remaining points on the 45-degree line (blue dots) are virtual fixed points of branch $f_M(x)$ of map $M$. In contrast to map $C$, steady-state mispricing is now bounded, with its maximal level given by $h$. This is ensured by the market entry behavior of fundamentalists. The Jacobian matrix of branch $f_M(x)$ of map $M$ is identical to that of map $C$, i.e., $J_C = J_M$.

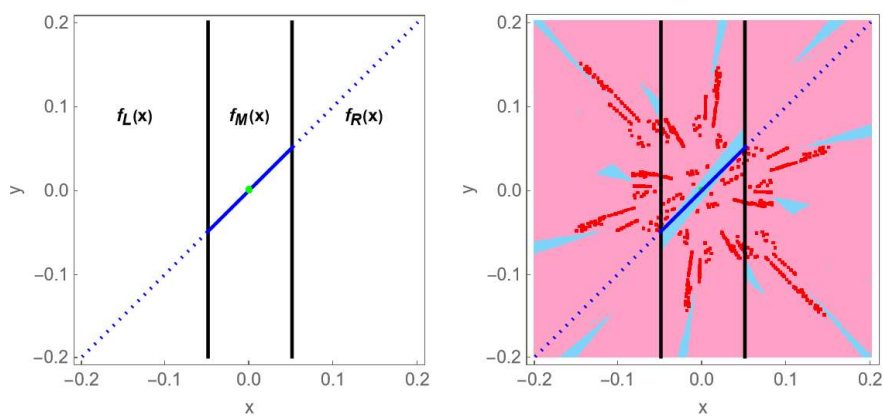

Figure 4: Properties of map $M$. The left panel shows the areas of definition for the branches $f_L(x)$, $f_M(x)$, and $f_R(x)$ of map $M$. The segment $S^*$ of fixed points and the origin are depicted in blue and green, respectively. The right panel shows the basin of attraction of the segment $S^*$ of fixed points in light blue. The initial conditions marked in light red converge to a weird quasiperiodic attractor, represented by the red dots. Parameter setting: $b = 0.80$, $c = 2.50$, and $h = 0.05$.

What might the dynamics of map $M$ look like? Figure 5, based on $b = 0.80$, $c = 2.50$, and $h = 0.05$, presents two examples. The left (right) panel portrays the dynamics in the time domain (state space). The first example starts with the initial conditions $x_0 = -0.12$ and $y_0 = -0.16$. After a few time steps, the stock price has converged to one of the nonfundamental fixed points of map $M$, yielding constant mispricing. From now on, we will depict such fixed-point dynamics in blue. The second example rests on the initial conditions $x_0 = 0.13$ and $y_0 = 0.00$. Here, the stock price exhibits endogenous dynamics, an outcome that we will now always report in red. While mispricing is bounded, it oscillates, giving rise to a weird quasiperiodic attractor.[3] Obviously, the initial conditions matter.

To explore the dependence of our model's dynamics on the initial conditions in more detail, we visualize the basins of attraction of map $M$'s coexisting attractors in the top right panel of Figure 4. Computations of basins of attraction in our paper are consistently conducted using a 250 by 250 grid. Here, the initial conditions are varied in the intervals $-0.20 < x < 0.20$ and $-0.20 < y < 0.20$. The basin of attraction of the segment of fixed points $S^*$ is marked in light blue, while all other initial conditions, represented in light red, converge to a weird quasiperiodic attractor, depicted in red. Weird quasiperiodic attractors are a completely new dynamical phenomenon, which we study in more detail in the sequel.

---

[3] Gardini et al. (2024c) show that the emergence of weird quasiperiodic attractors does not hinge on the assumption that all fundamentalists are risk averse, respectively, that map $M$ possesses a continuum of nonfundamental fixed points.

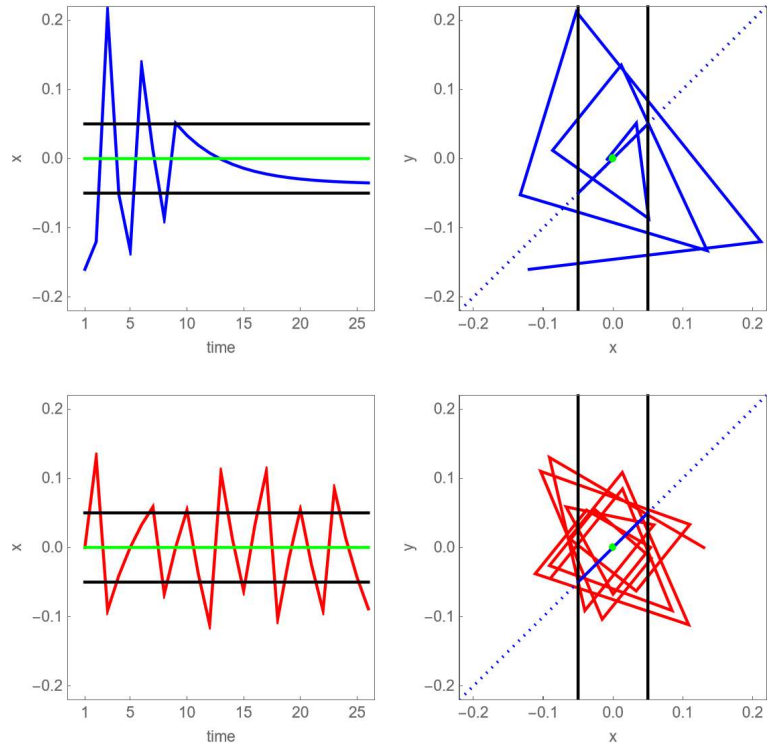

Figure 5: Dynamics of map $M$ in the time domain and in state space for $b = 0.80$, $c = 1.35$, and $h = 0.05$. Initial conditions: $x_0 = -0.12$ and $y_0 = -0.16$ (top) and $x_0 = 0.13$ and $y_0 = 0.00$ (bottom).

The dynamics of our model naturally depend on the underlying parameter settings. To broaden our perspective, Figure 6 presents two-dimensional bifurcation diagrams using a 250 by 250 grid, where we vary parameters $b$ and $c$ within the ranges $0.00 < b < 1.10$ and $0.00 < c < 4.40$, while setting parameter $h$ to $h = 0.05$. In the left panel, the initial conditions are $x_0 = 0.06$ and $y_0 = 0.06$. Green dots indicate convergence to the fundamental value, blue dots represent convergence to a nonfundamental value, red dots imply endogenous dynamics, and gray dots reflect divergent dynamics. Notably, no green dots are present – none of the simulations resulted in convergence to the fundamental value. The right panel repeats this exercise with the initial conditions $x_0 = 0.04$ and $y_0 = 0.00$. For low values of parameter $b$, previously divergent dynamics transition into fixed-point dynamics. Our model therefore contradicts Fama's (1970) claim that financial markets are informationally efficient.

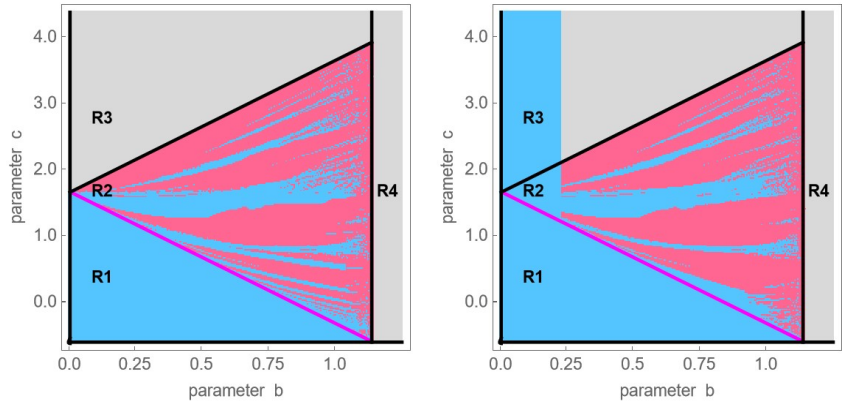

Figure 6: Two-dimensional bifurcation diagrams for map $M$. Parameters $b$ and $c$ are varied between $0.00 < b < 1.10$ and $0.00 < c < 4.40$, while parameter $h$ is set to $h = 0.05$. Initial conditions: $x_0 = 0.06$ and $y_0 = 0.06$ (left) and $x_0 = 0.04$ and $y_0 = 0.00$ (right). Blue dots: fixed point dynamics. Red dots: weird quasiperiodic dynamics. Gray dots: divergent dynamics.

It is insightful to compare these numerically generated findings with those reported in Figure 1. The parameter space appears to be divided into four regions, displaying (1) convergence to a nonfundamental fixed point, (2) convergence to a nonfundamental fixed point or the emergence of endogenous dynamics, (3) convergence to a nonfundamental fixed point or divergent dynamics, and (4) divergent dynamics. Apart from the magenta line, the boundaries of these regions are already visible in Figure 1. The next section reveals that we can rigorously prove this hypothesis, identify the magenta boundary, characterize the basins of attraction of the coexisting attractors, and formally establish the appearance of weird quasiperiodic attractors.

### 3.2 Main results

Propositions 1 and 2, proven in Appendix A, present our main analytical results. Proposition 1 characterizes the basins of attraction of the nonfundamental fixed points, while Proposition 2 reveals the types of dynamics our model may produce with respect to the market impact of chartists and fundamentalists.

**Proposition 1:** *For $0 < b < 1$, the basin of attraction of the segment $S^*$ of fixed points consists of an immediate basin $\mathcal{B}_0(S^*)$, and all its preimages of any rank, where $\mathcal{B}_0(S^*)$ is a parallelogram with vertices $(-h, -h)$, $(h, -h + 2h/b)$, $(h, h)$, and $(-h, h - 2h/b)$.*

Note that the boundaries of the immediate basin of attraction of the segment $S^*$ of fixed points depend solely on parameters $b$ and $h$. Why is this the case? Since fundamentalists are inactive when the stock price lies between $-h$ and $h$, the market impact of fundamentalists, represented by parameter $c$, does not influence $\mathcal{B}_0(S^*)$. However, the size of $\mathcal{B}_0(S^*)$ increases in line with parameter $h$ and decreases in line with parameter $b$.

We discuss these and some further implications of Proposition 1 using a few examples. The top line of panels of Figure 7 is based on $b = 0.80$, $c = 1.00$, and $h = 0.10$. The light blue dots in the left panel represent the initial conditions that converge to a nonfundamental fixed point, while the light red dots converge to a weird quasiperiodic attractor, represented in red. While these basins of attraction are identified numerically, we now have analytical insights about the boundaries of the immediate basin of attraction of the segment $S^*$ of fixed points, marked in yellow. The right panel shows the dynamics of the weird quasiperiodic attractor in the time domain.

Suppose policymakers are able to reduce parameter $h$, for example, by implementing regulations that decrease the risks associated with stock market trading. The middle line of panels of Figure 7, based on $b = 0.80$, $c = 1.00$, and $h = 0.05$, illustrates the effects of such a policy. Recall that parameter $h$ is a scaling factor. By reducing parameter $h$ from $h = 0.10$ to $h = 0.05$, the magnitude of stock price fluctuations associated with the weird quasiperiodic attractor is halved. The same is true for the segment $S^*$ of fixed points, leading to reduced mispricing when the stock price convergence towards a nonfundamental fixed point, and also its immediate basin of attraction shrinks.

Alternatively, suppose policymakers are able to reduce parameter $b$, for instance, by implementing regulations that diminish the market impact of chartists. The bottom line of panels of Figure 7, based on $b = 0.60$, $c = 1.00$ and $h = 0.05$, illustrates the effects of such a policy. As shown, the size of the immediate basin of attraction of the segment $S^*$ of fixed points expands. This policy has no impact on the segment $S^*$ of fixed points, and the magnitude of price fluctuations remains relatively stable.

Figure 7 also reveals that the total basin of attraction for the two coexisting attractors changes in a nontrivial way, highlighting the complexity of our model's behavior. We will see more examples of this when we discuss the implications of the following proposition.

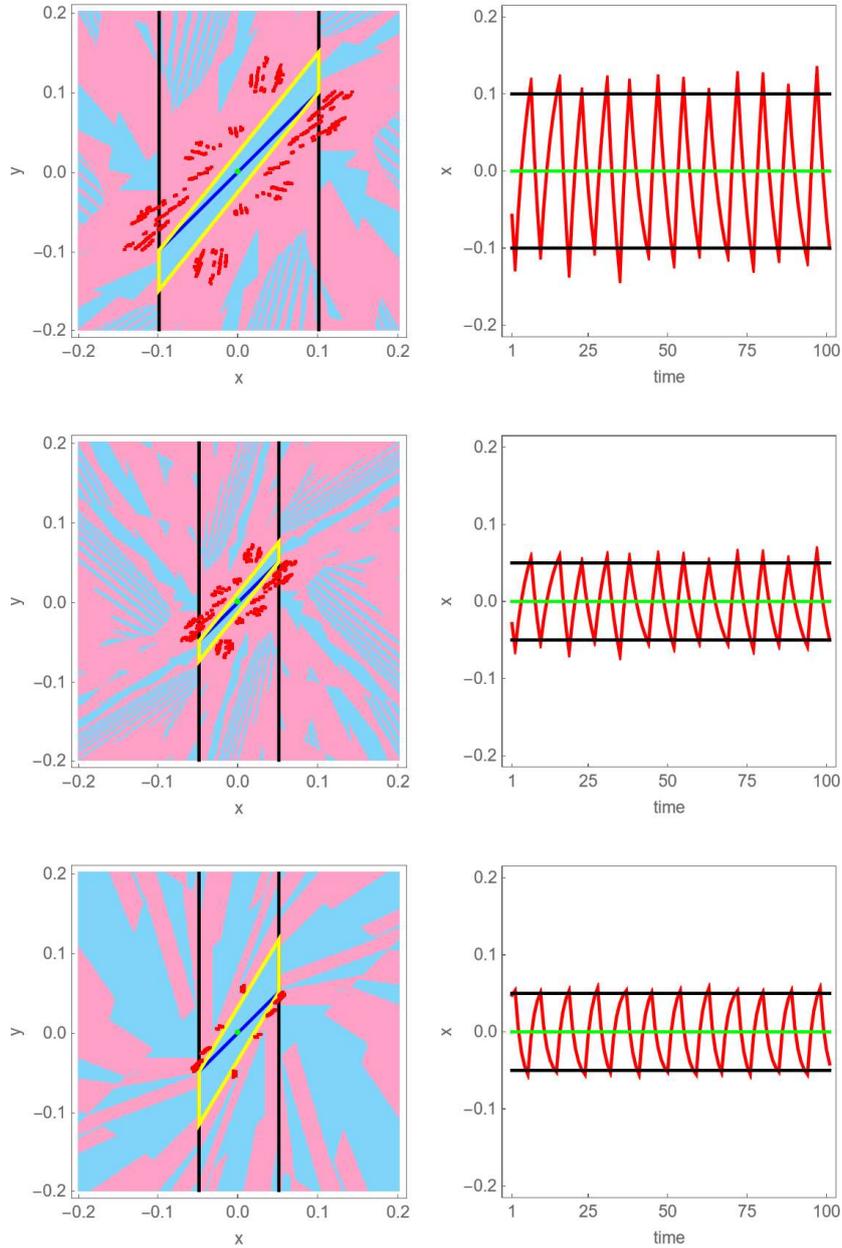

Figure 7: Immediate basin of attraction of the segment $S^*$ of fixed points. Left: The boundaries of the immediate basin of attraction of the segment $S^*$ of fixed points are marked in yellow. The segment $S^*$ of fixed points is represented in blue, and the origin is shown in green. The initial conditions, depicted in light blue and light red, converge to nonfundamental fixed points and a weird quasiperiodic attractor (shown in red), respectively. Right: Time series dynamics of the weird quasiperiodic attractor. Parameter setting: $b = 0.80$, $c = 1.00$, and $h = 0.10$ (top); $b = 0.80$, $c = 1.00$, and $h = 0.05$ (middle); and $b = 0.60$, $c = 1.00$, and $h = 0.05$ (bottom).

**Proposition 2:** *(1) For $b > 1$, all trajectories are divergent except for the segment $S^*$ of saddle fixed points. (2) For $0 < b < 1$ and $c > 2(1 + b)$, the basin of attraction of the segment $S^*$ of fixed points is given by $\mathcal{B}_0(S^*)$ and a sequence of disconnected preimages of any rank of this set via the inverses $f_L^{-1}$ and $f_R^{-1}$; all other points have divergent trajectories. (3) For $0 < b < 1$ and $c < 2(1 - b)$, the segment $S^*$ of fixed points is globally attracting. (4) For $0 < b < 1$ and $2(1 - b) < c < 2(1 + b)$, weird quasiperiodic attractors may coexist with the attracting segment $S^*$ of fixed points.*

According to Proposition 2, there are indeed four parameter regions, see again Figure 6, that lead to different outcomes. In region R1, given by $0 < b < 1$ and $c < 2(1 - b)$, the market impact of both chartists and fundamentalists is relatively low. As a result, the generic trajectory of the stock price converges to a nonfundamental fixed point.[4] In region R2, given by $0 < b < 1$ and $2(1 - b) < c < 2(1 + b)$, the market impact of chartists and fundamentalists is moderate. Consequently, the generic trajectory of the stock price either converges to a nonfundamental fixed point or exhibits endogenous dynamics.[5] In region R3, given by $0 < b < 1$ and $c > 2(1 + b)$, the market impact of fundamentalists is relatively high. In this region, the generic trajectory of the stock price either converges to a nonfundamental fixed point or displays divergent dynamics. In region R4, given by $b > 1$, the market impact of chartists is relatively high, and the generic trajectory of the stock price follows a divergent path.

Figures 8 to 11 provide a selection of examples that further illustrate the dynamics of map $M$ and the implication of Propositions 1 and 2. Figures 8 and 9 show basins of attraction and time series dynamics for increasing values of parameter $c$, while keeping parameters $b$ and $h$ equal to $b = 0.80$ and $h = 0.05$. For $c = 0.25$, the generic trajectory of the stock price converges to a nonfundamental fixed point located on segment $S^*$. In fact, the typical behavior in region R1 is that the stock market displays permanent constant mispricing. For $c$ equal to 1.00, 1.70, 2.05, and 2.50, the generic trajectory of the stock price either converges to a nonfundamental fixed point or it displays weird quasiperiodic dynamics. All these examples are associated with region R2.

---

[4] Only those points that lie on the line $y = x/b$ within the strip between $-h$ and $h$, as well as their preimages, converge to the fundamental value.

[5] Remarkably, our model predicts that the stock price may oscillate around its fundamental value for parameter combinations that are usually associated with stable dynamics. For the linear chartist-fundamentalist model, i.e., map $F$, this is not the case. In particular, the green parameter space depicted in the left panel of Figure 1 yields ordinary fixed point dynamics.

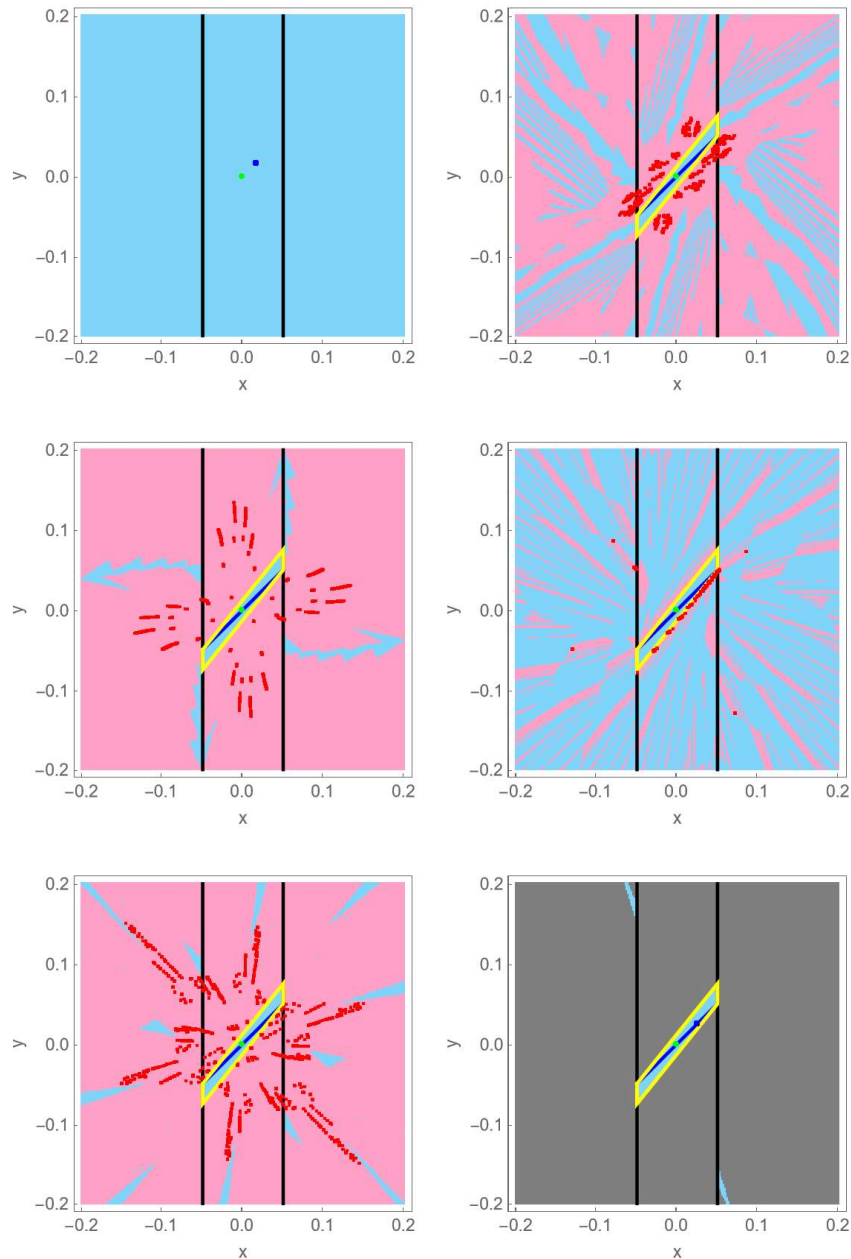

Figure 8: Basins of attraction for different values of parameter $c$. From top left to bottom right, parameter $c$ takes the values 0.25, 1.00, 1.70, 2.05, 2.50, and 3.61, while the remaining parameters are set to $b = 0.80$ and $h = 0.05$. The initial conditions taken from the light blue, light red, and gray area result in fixed point, weird quasiperiodic, and divergent dynamics, respectively. Fixed point and weird quasiperiodic attractors are superimposed in blue and red, respectively.

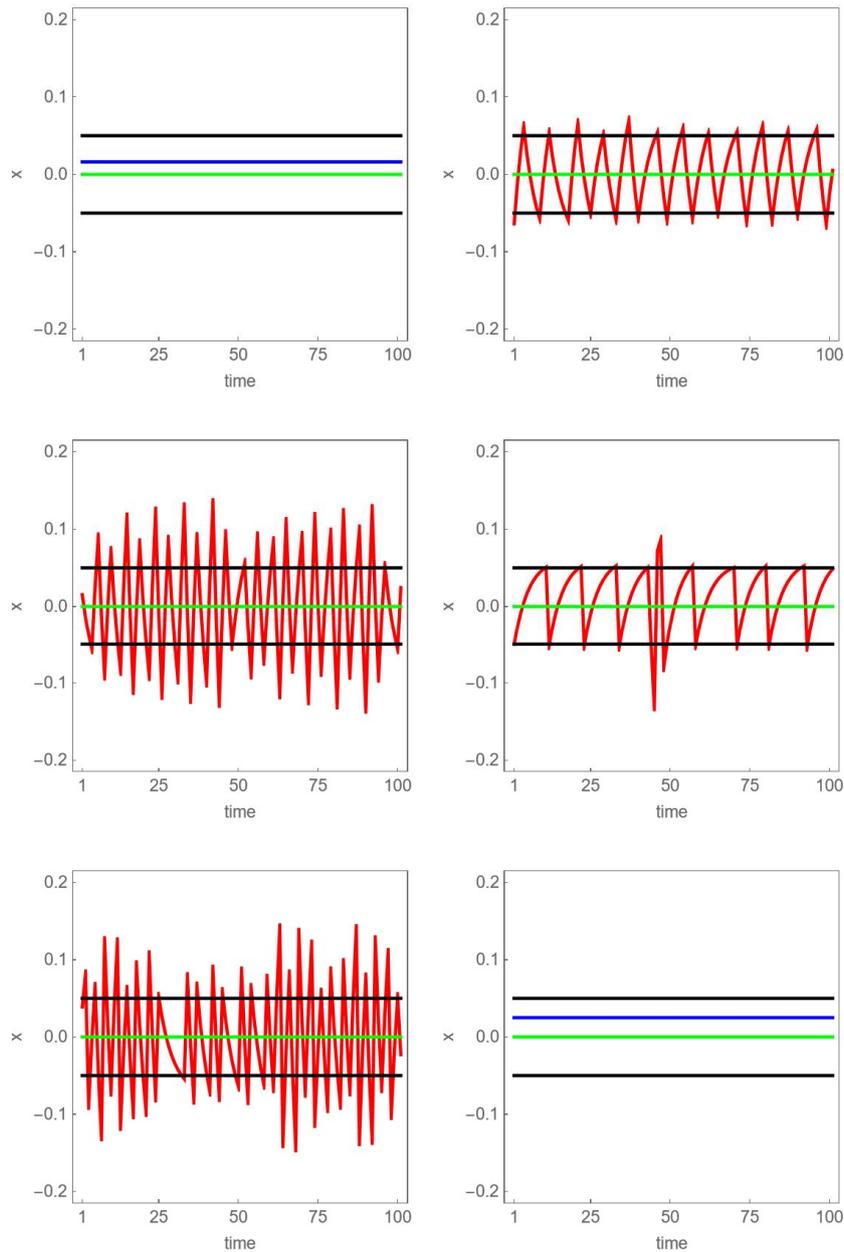

Figure 9: Dynamics in the time domain for different values of parameter $c$. The same parameter setting as in Figure 8. Green line: fundamental value. Blue line: fixed point dynamics. Red line: weird quasiperiodic dynamics. Black lines: $x = \pm h$.

While the basins of attraction that give rise to weird quasiperiodic attractor tend to increase with parameter $c$, there are exceptions. For instance, at $c = 2.05$, the majority of the initial conditions result in constant mispricing. In addition, the initial conditions that lead to endogenous dynamics produce a weird quasiperiodic attractor that is not symmetric with respect to the origin, revealing the existence of a second weird periodic attractor that is symmetric to it with respect to the origin. See Appendix B for a deeper discussion. At $c = 3.61$, the generic trajectory of the stock price either converges to a nonfundamental fixed point or diverges. Once again, note that the immediate basin of attraction of the segment $S^*$ of fixed points is independent of parameter $c$.

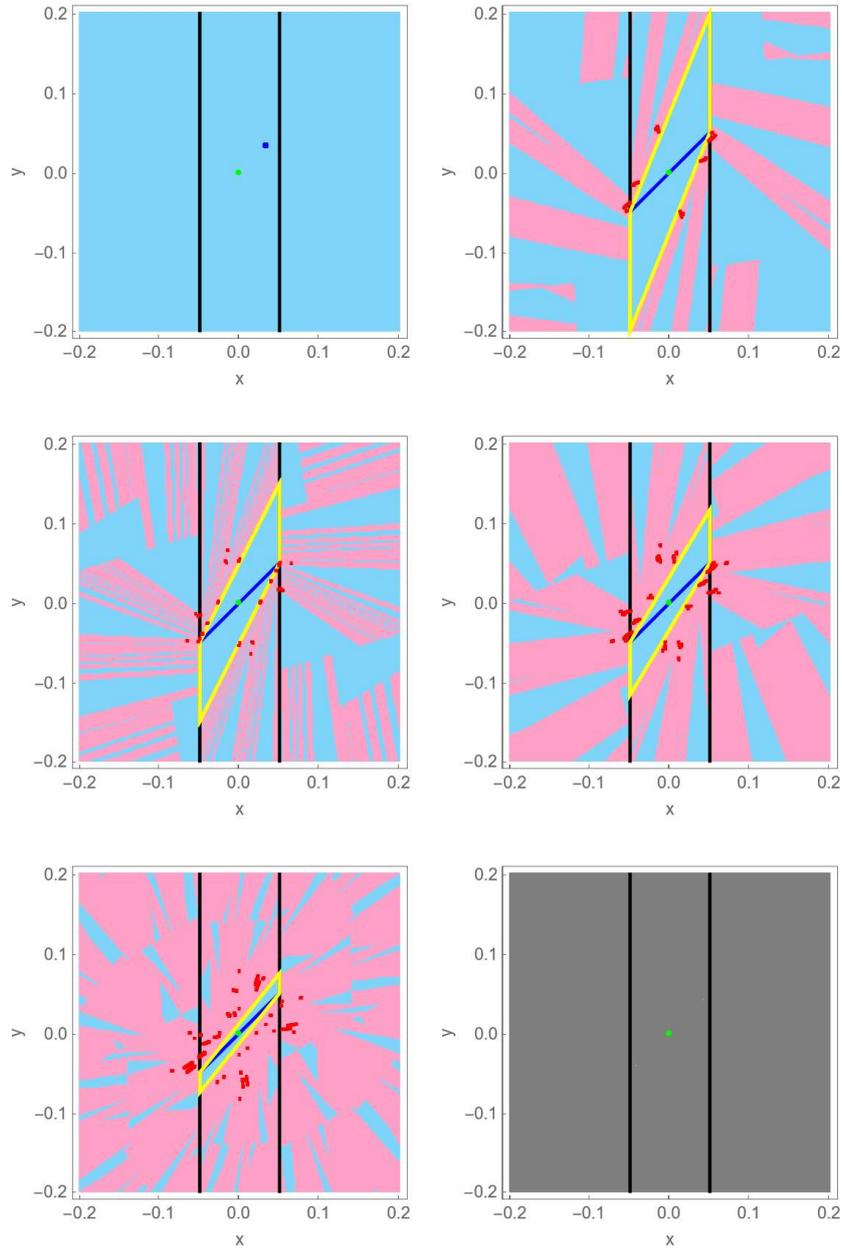

Figure 10: Basins of attraction for different values of parameter $b$. From top left to bottom right, parameter $b$ takes the values 0.30, 0.40, 0.50, 0.60, 0.80, and 1.05, while the remaining parameters are set to $c = 1.35$ and $h = 0.05$. The initial conditions taken from the light blue, light red, and gray area result in fixed point, weird quasiperiodic, and divergent dynamics, respectively. Fixed point and weird quasiperiodic attractors are superimposed in blue and red, respectively.

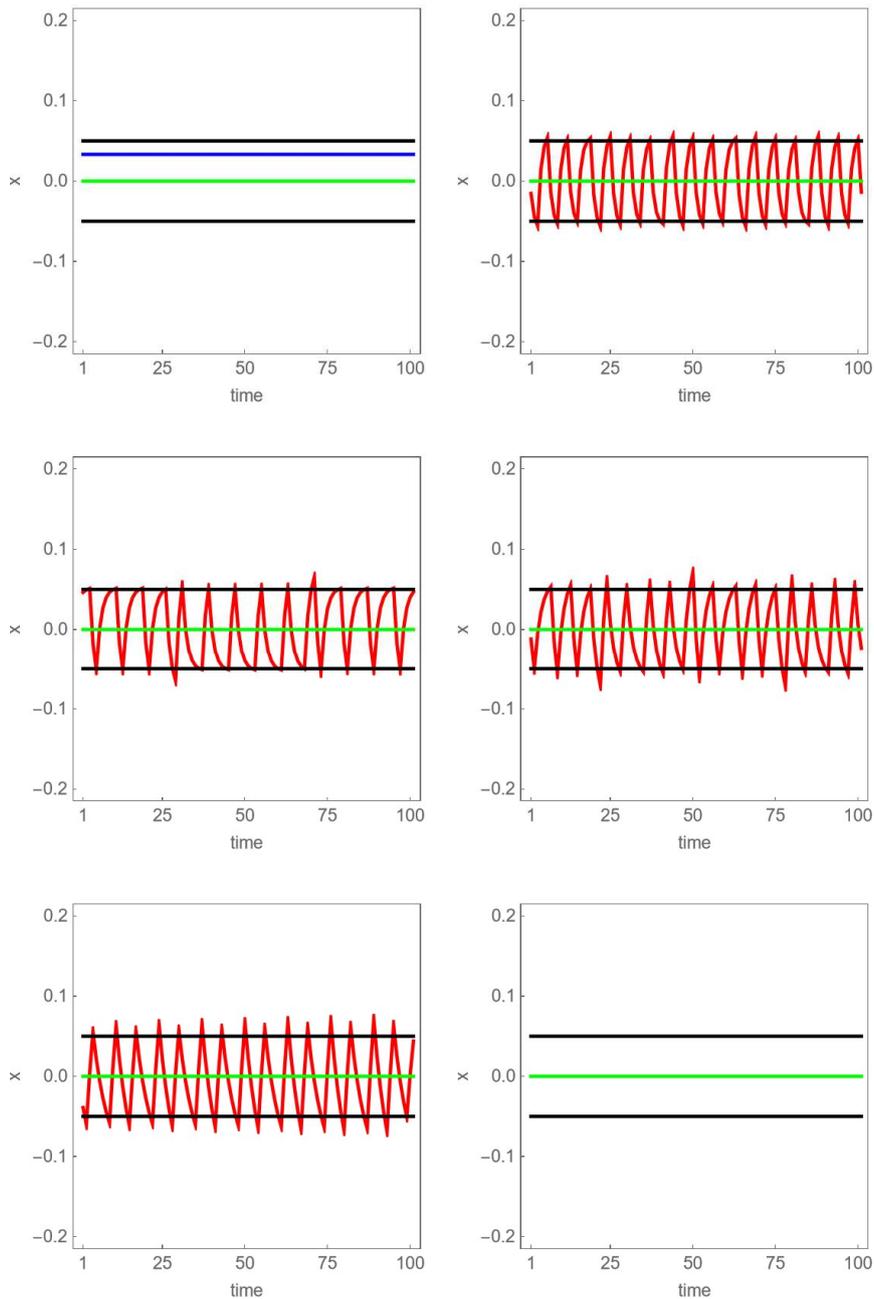

Figure 11: Dynamics in the time domain for different values of parameter $b$. The same parameter setting as in Figure 10. Green line: fundamental value. Blue line: fixed point dynamics. Red line: weird quasiperiodic dynamics. Black lines: $x = \pm h$.

Figures 10 and 11 show basins of attraction and time series dynamics for increasing values of parameter $b$, while keeping parameters $c$ and $h$ equal to $c = 1.35$ and $h = 0.05$. For $b = 0.30$, the generic trajectory of the stock price converges to a nonfundamental fixed point. In region R1, the stock market always displays constant mispricing. For $b$ equal to 0.40, 0.50, 0.60, and 0.80, the generic trajectory of the stock price either converges to a nonfundamental fixed point or displays weird quasiperiodic dynamics, as to be expected in region R2. At $b = 1.05$, the stock market is subject to divergent dynamics, except when the initial conditions are located on the segment $S^*$ of fixed points. See part (1) of Proposition 2. Note that the immediate basin of attraction of the segment $S^*$ of fixed points shrinks with parameter $b$, until it eventually disappears.

## 3.3 The effects of fundamental shocks

Figure 12 illustrates examples where our model's dynamics are subject to fundamental exogenous shocks. From top to bottom, we set parameter $c$ to $0.45$, $0.75$, and $1.10$, respectively, while the other parameters are fixed at $b = 0.80$, $h = 0.05$, and $\sigma_d = 0.005$. All three parameter constellations fall within region R2. The evolution of the stochastic stock price is marked in purple. The left panels show the dynamics in the time domain and the right panels in state space. Of course, the computations of the basins of attraction displayed in the right panels of Figure 12 are based on our deterministic stock market model.

As can be seen, stock prices alternate between the basin of attraction of the segment $S^*$ of fixed points and the basin of attraction of the weird quasiperiodic attractor. We understand the dynamics as follows. When stock prices are within the basin of attraction of the segment $S^*$ of fixed points, the stock market tends to experience mild fluctuations. Due to the fundamental shocks, however, the stock price eventually exits this basin. For a time, the stock market then undergoes significant fluctuations until it returns to the previous basin. Importantly, when stock prices are in the basin of attraction of the weird quasiperiodic attractor, a substantial part of the stock market's dynamics has an endogenous nature.

Note that the size and structure of the basins of attraction influence the duration of the two regimes and the frequency of regime shifts. For instance, as parameter $c$ increases from $0.45$ to $1.10$, the total basin of attraction of the weird quasiperiodic attractor expands, making the volatile regime more dominant. Similar dynamics are observed with increasing values of parameter $b$.

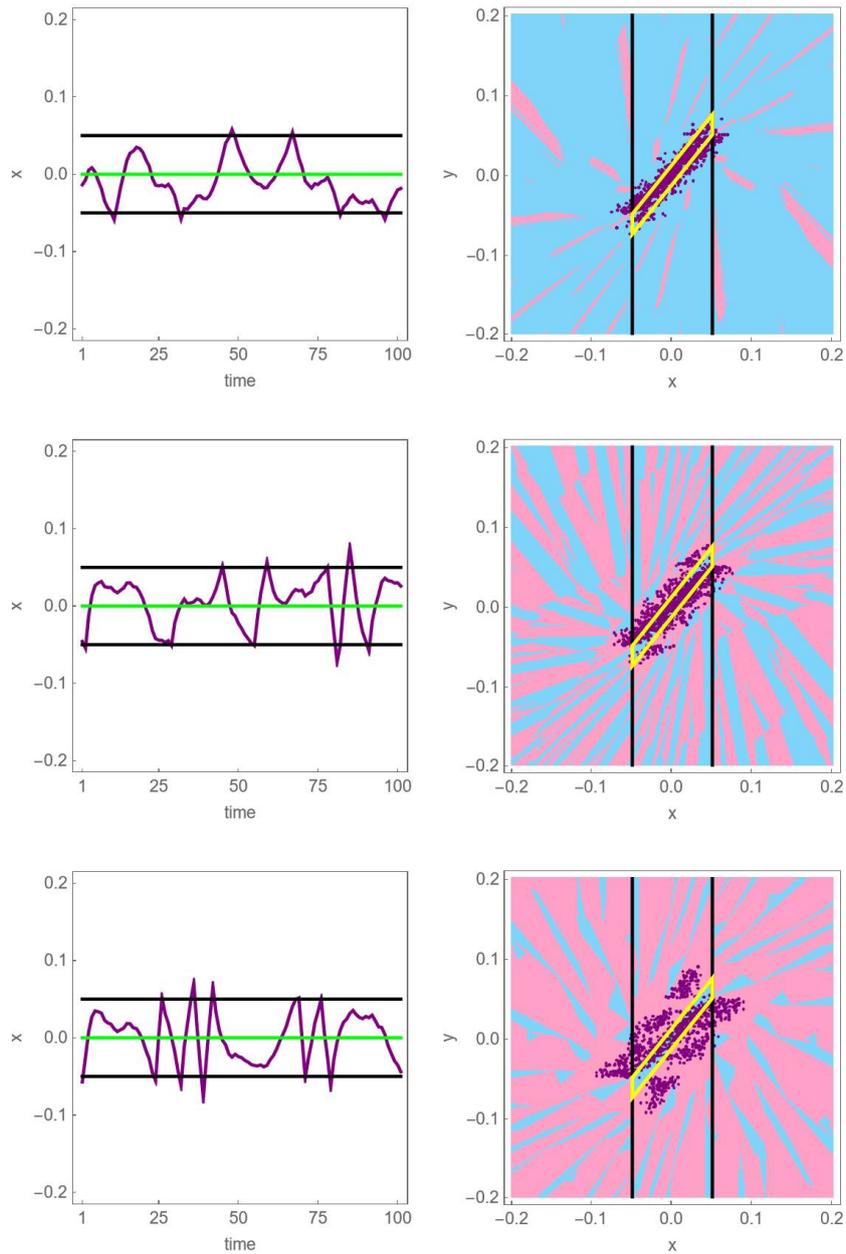

Figure 12: Stochastic stock market dynamics. From top to bottom, parameter $c$ is set to 0.45, 0.75, and 1.10, respectively, with parameters $b = 0.80$, $h = 0.05$, and $\sigma_d = 0.005$ held constant.

# 4 Conclusions

We examine the extent to which stock markets are informationally efficient using a chartist-fundamentalist model. Fama (1970) argues that stock prices reflect all available information. As demonstrated by Samuelson (1965), stock prices then fluctuate randomly. Grossman and Stiglitz (1980) emphasize that speculators need an incentive to gather, analyze, and act on information, which leads to constant mispricing. According to Lo and Farmer (1999), a stock market's mispricing should be seen as dynamic rather than static, as speculators are boundedly rational and regularly adjust their behavior in response to their market experiences. Our model reveals the possibility of coexistence between constant and oscillatory mispricing, reconciling the views of both Grossman and Stiglitz (1980) and Lo and Farmer (1999). In addition, we have identified the existence of a weird quasiperiodic attractor – a novel dynamic behavior that blurs the lines between quasiperiodic and chaotic dynamics, deepening our understanding of the complex behavior of stock markets. Overall, this discovery underscores the importance of considering piecewise-linear discontinuous models in the analysis of economic systems.

**Acknowledgements**

Laura Gardini thanks the VSB-Technical University of Ostrava (SGS Research Project SP2024/047) and the European Union (REFRESH Project-Research Excellence for Region Sustainability and High-Tech Industries of the European Just Transition Fund, Grant CZ.10.03.01/00/22 003/000004). The work of Davide Radi and Iryna Sushko was funded by PRIN 2022 under the Italian Ministry of University and Research (MUR) Prot. 2022JRY7EF - Qnt4Green - Quantitative Approaches for Green Bond Market: Risk Assessment, Agency Problems and Policy Incentives. Davide Radi also acknowledges financial support from the Gruppo Nazionale di Fisica Matematica GNFM-INdAM.

**Appendix A**

Proof of Proposition 1:

The eigenvalues of the Jacobian matrix $J_M$, associated with the middle partition of map $M$, are $\lambda_1 = 1$, with eigenvector $r_1 = (1,1)$ along the main diagonal, and $\lambda_2 = b$, with eigenvector $r_2 = (1,1/b)$. It follows that for $0 < b < 1$, the attracting eigenvector originating from each fixed point $(u,u)$ of the segment $S^*$ of fixed points is a segment of the straight line described by the equation $y = u + (x - u)/b$. This segment must belong to the middle partition of map $M$, i.e., for $-h <$ $x < h$. From the point $(-h, -h)$, only the upper segment exits, corresponding to the line $y = -h + (x + h)/b$, which intersects the discontinuity line $x = h$ at the point $(h, -h + 2h/b)$. This defines the upper boundary of the immediate basin of attraction of the segment $S^*$ of fixed points, denoted by $\mathcal{B}_0(S^*)$. Conversely, from the point $(h, h)$, only the lower segment exits, corresponding to the line $y = h + (x - h)/b$, which intersects the discontinuity line $x = -h$ at the point $(-h, h - 2h/b)$. This defines the lower boundary of $\mathcal{B}_0(S^*)$. ∎

A few comments are in order. The first preimages of the immediate basin of attraction of the segment $S^*$ of fixed points, i.e., $f_L^{-1}(\mathcal{B}_0(S^*))$ and $f_R^{-1}(\mathcal{B}_0(S^*))$, consist of two symmetric triangles, each with a segment on the discontinuity lines $x = -h$ and $x = h$, respectively, via $f_{L/R}^{-1}(u, v) = (v, \frac{(1+b-c)v-u}{b})$. From $f_L^{-1}$, we obtain the vertices of the triangle in the left partition. They are given by

(A) $f_L^{-1}(-h, -h) = (-h, h\frac{c-b}{b})$

(B) $f_L^{-1}(-h, -h(\frac{2}{b} - 1)) = (-h(\frac{2}{b} - 1), \frac{h}{b}(-1 + (1 + b - c)(\frac{2}{b} - 1)))$

(C) $f_L^{-1}(h(1 - 2b), -h) = (-h, \frac{h}{b}(b + c - 2))$

The triangle in the left partition of map $M$ is the preimage of the triangle of $\mathcal{B}_0(S^*)$ in the region $y < -h$. The triangle in the right partition of map $M$ is the preimage of the triangle of $\mathcal{B}_0(S^*)$ in the region $y > h$. It follows that no point from the external partitions of map $M$ can be mapped in one iteration in the strip of $\mathcal{B}_0(S^*)$ with $-h < y < h$, which includes the segment of fixed points $S^*$. This portion has no rank-1 preimage.

Figure A1, based on $b = 0.4$, $c = 2.85$, and $h = 0.05$, provides an example. White points converge towards the segment $S^*$ of fixed points, while gray points exhibit divergent dynamics. The left panel highlights the boundaries of the immediate basin of attraction of the segment of fixed points $S^*$ in yellow. The right panel shows that points $A$, $B$, and $C$ are mapped to points $A'$, $B'$, and $C'$, respectively.

Proof of part (1) of Proposition 2:

For $b > 1$, the eigenvector originating from each fixed point $(u, u)$ of the segment $S^*$ of fixed points is repelling (local unstable set), and its points exit from the middle partition of map $M$ after a finite number of iterations. Since the dynamics in the external partitions of map $M$ are also expanding, with real or complex conjugate eigenvalues outside the unit circle, the generic trajectory is divergent. The only points with non-divergent trajectories are those on the segment $S^*$ of saddle fixed points, and such points have no rank-1 preimage under $f_{L,R}^{-1}$. ∎

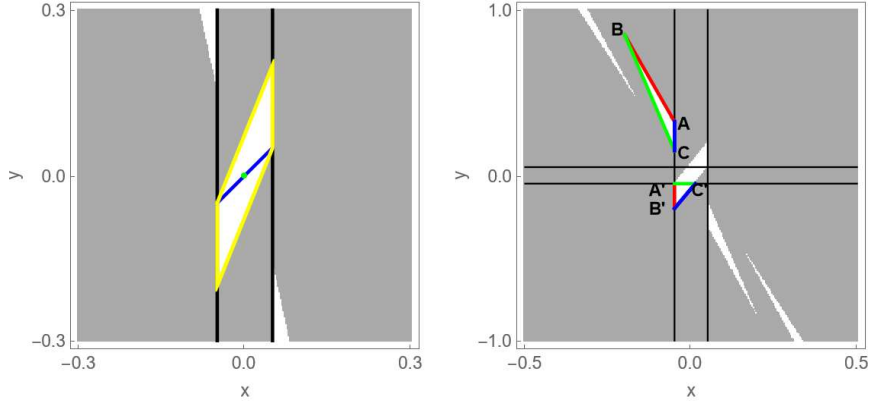

Figure A1: The basin of attraction of the segment $S^*$ of fixed points. White points converge towards the segment $S^*$ of fixed points, while gray points exhibit divergent dynamics. Parameter setting: $b = 0.40$, $c = 2.85$, and $h = 0.05$.

Proof of part (2) of Proposition 2:

For $0 < b < 1$ and $c > 2(1 + b)$, the two functions $f_{L/R}$ of map $M$ have a virtual saddle fixed point at the origin, with two real negative eigenvalues $\lambda_i = \frac{1}{2}((1 + b - c) \pm \sqrt{(1 + b - c)^2 - 4b})$, $i = 1,2$. The eigenvectors associated with these eigenvalues lie along the lines $y = x/\lambda_i$. Therefore, in the external partitions of map $M$, points not belonging to the basin of attraction of the segment $S^*$ of fixed points have divergent trajectories.

From the first preimages of $\mathcal{B}_0(S^*)$, specifically $f_L^{-1}(\mathcal{B}_0(S^*))$, point (C), introduced after the proof of Proposition 1, lies above the immediate basin of attraction of the segment $S^*$ of fixed points, as $\frac{h}{b}(b + c - 2) > h$ (which occurs when $c > 2$). Thus, the triangle $f_L^{-1}(\mathcal{B}_0(S^*))$ is disjoint from $\mathcal{B}_0(S^*)$, and the total basin of attraction of the segment of fixed points $S^*$ consists of disconnected elements. ∎

Proof of part (3) of Proposition 2:

Let us first assume that the eigenvalues of the Jacobian matrix $J_{L/R}$ are real and positive. In this case, the points in the middle partition of map $M$ that lie outside the immediate basin of attraction of the segment $S^*$ of fixed points have trajectories that enter one of the two external partitions of map $M$ within a finite number of steps. Points from the external partitions of map $M$ are attracted to the virtual attracting fixed point at the origin, so that any trajectory eventually enters $\mathcal{B}_0(S^*)$ in a finite number of steps.

Before we consider the cases where the eigenvalues of the Jacobian matrix $J_{L/R}$ are real and negative or complex conjugate, note that the boundary $c = 2(1-b)$ is related to the geometric shape of the rank-1 preimage of $\mathcal{B}_0(S^*)$ under $f_L^{-1}$. The following results hold. The image of $A' = (-h, -h)$ via $f_L$ lies exactly on the lower boundary of the immediate basin of attraction of the segment $S^*$ of fixed points when $A'' = f_L(-h, -h) = (-h(1-c), -h)$ lies on the line described by the equation $y = h + (x-h)/b$, which occurs only for $c = 2(1-b)$.

This means that for $c < 2(1-b)$, the point $A'' = f_L(-h, -h)$ is inside the immediate basin of attraction of the segment $S^*$ of fixed points. Furthermore, a segment of the left boundary of $\mathcal{B}_0(S^*)$ is mapped via both $f_L$ and $f_M$ into the immediate basin of attraction of the segment $S^*$ of fixed points. See the left and middle panels of Figure A2. In fact, the triangle preimage of the immediate basin of attraction of the segment $S^*$ of fixed points via $f_L^{-1}$ includes a segment of $x = -h$ that overlaps with $\mathcal{B}_0(S^*)$.

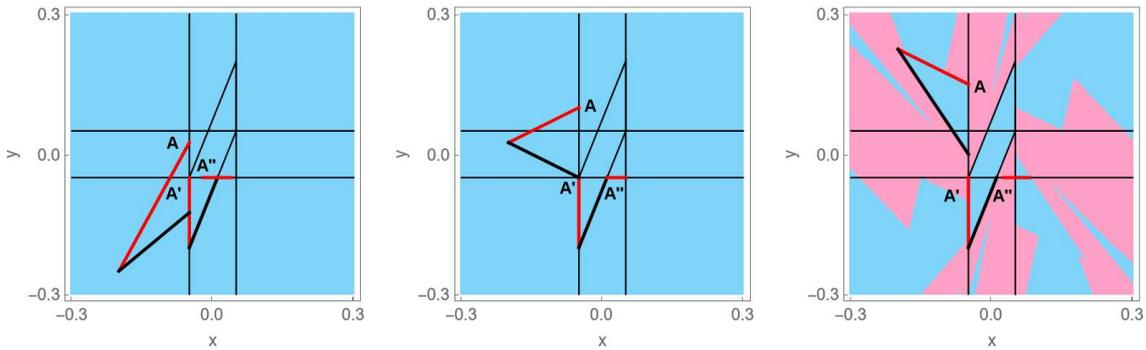

Figure A2: Basin of attraction of segment $S^*$ of fixed points. Left: $b = 0.40$, $c = 0.60$, and $h = 0.05$. Center: $b = 0.40$, $c = 1.20$, and $h = 0.05$. Right: $b = 0.40$, $c = 1.60$, and $h = 0.05$.

When the eigenvalues of $J_{L/R}$ are complex conjugate, a point in the external partitions of map $M$ is attracted to the virtual fixed point at the origin, rotating around it with the trajectory of points forming suitable arcs of spirals. A similar behavior also occurs when the eigenvalues of $J_{L/R}$ are real and negative. In this case, points outside the middle partition of map $M$ flip between its left and right partition, with trajectories that tend toward the origin.

In both cases, it follows that within a finite number of iterations, a point of the trajectory will enter $\mathcal{B}_0(S^*)$ or one of its rank-1 symmetric preimages $f_{L/R}^{-1}(\mathcal{B}_0(S^*))$. In other words, since $\mathcal{B}_0(S^*) \cup f_L^{-1}(\mathcal{B}_0(S^*)) \cup f_R^{-1}(\mathcal{B}_0(S^*))$ forms a connected set, also the total basin of attraction of the segment $S^*$ of fixed points is a connected set, encompassing the entire phase plane. Thus, the segment $S^*$ of fixed points is globally attracting. ∎

Proof of part (4) of Proposition 1:

When $c > 2(1-b)$, the point $A'' = f_L(-h, -h)$ lies outside the immediate basin of attraction of the segment $S^*$ of fixed points. As a result, the rank-1 symmetric preimages $f_{L/R}^{-1}(\mathcal{B}_0(S^*))$ are disjoint from $\mathcal{B}_0(S^*)$ and the total basin of attraction of the segment $S^*$ of fixed points may be disconnected. This leads to the possible existence of points with the trajectories that cross all three partitions of map $M$ but never enter $\mathcal{B}_0(S^*)$. Some points of the line $x = -h$ between the points $C$ and $A'$ may belong to the basin boundary of a coexisting attractor.

When there exists an attracting set different from the segment $S^*$ of fixed points, numerical evidence shows that the trajectory of $A'' = f_L(-h, -h)$ converges to that attractor. However, if its trajectory enters $\mathcal{B}_0(S^*)$, then the segment $S^*$ of fixed points is globally attracting.

Recall that divergent trajectories cannot exist in this parameter range. Regarding the structure of the attractor that coexists with the segment $S^*$ of fixed points, consider a cycle of period $k$ with symbolic sequence $\sigma_1 \ldots \sigma_k$, where $\sigma_k \in \{L, M, R\}$. Then $M^k(X) = J_k X$, with $J_k = J_{\sigma_k} \ldots J_{\sigma_1}$. A periodic point $X$ of the cycle must satisfy the equation $J_k X = X$.

Two cases are possible:

(i) If $J_k$ does not have an eigenvalue of $+1$, then the only solution is $X = 0$, and such a $k$-cycle cannot exist.

(jj) If $J_k$ has an eigenvalue of $+1$, then it is possible to have $k$ segments filled with fixed points of map $M^k$ (corresponding to points of a $k$-cycle), which are stable but not attracting for parameters inside the stability region of map $F$. However, the segments are attracting, since $\det(J_k) = b^k < 1$.

Clearly, case (i) is the generic case. The attracting sets that coexist with $S^*$, which we call weird quasiperiodic attractor $\mathcal{A}$, have a "weird" geometric shape. These attracting sets cannot include any cycle and are clearly nonchaotic. The latter is due to the fact that in a chaotic set, the unstable cycles are dense and homoclinic, yet there are no cycles in $\mathcal{A}$. For any point $X$ of the attracting set $\mathcal{A}$, $M^n(X) = J_n X$ belongs to the bounded invariant set for any $n$ and point $X$ never becomes periodic, since $J_n X = X$ cannot occur.

Formally, we can characterize the attractor $\mathcal{A}$ coexisting with $S^*$ as follows. Let $\mathcal{B}$ be the total basin of attraction of $S^*$, and $P$ the complementary set $P = \mathcal{R}^2 \setminus \mathcal{B}$. Then $\mathcal{A} \subseteq \bigcap_{n \geq 0} M^n(P)$. This shows that when $\mathcal{B}$ is not the entire phase plane, another invariant attracting set must exist. ∎

**Appendix B**

Due to the symmetry of map $M$, any existing weird quasiperiodic attractor is either symmetric with respect to the origin, or one more weird quasiperiodic attractor symmetric with respect to the origin exists. Figure B1 provides an example. The left panel is based on $b = 0.40$, $c = 2.10$, and $h = 0.05$. Light blue points converge to the segment $S^*$ of fixed points, while light red and light brown points converge to the weird quasiperiodic attractors $\mathcal{A}$ and $-\mathcal{A}$, respectively. The appearance or disappearance of coexisting weird quasiperiodic attractors is associated with contact bifurcations. When two weird quasiperiodic attractors coexist, as a parameter is varied, they can merge due to a contact with their related basins of attraction, leading to a unique weird quasiperiodic attractor. The right panel is based on $b = 0.40$, $c = 2.15$, and $h = 0.05$. After the merging of the two weird quasiperiod attractors, a unique weird quasiperiodic attractor exists.

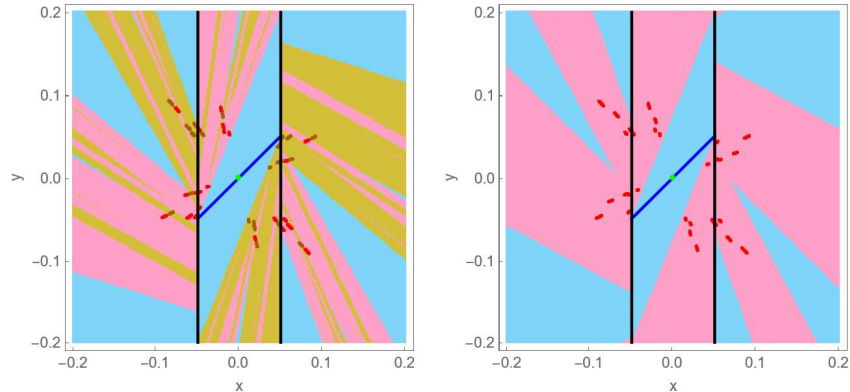

Figure B1: Appearance and disappearance of coexisting weird quasiperiodic attractors. The left panel is based on $b = 0.40$, $c = 2.10$, and $h = 0.05$. Light blue points converge to the segment of fixed points $S^*$, while light red and light brown points converge to the weird quasiperiodic attractor $\mathcal{A}$ and $-\mathcal{A}$, respectively. The right panel is based on $b = 0.40$, $c = 2.15$, and $h = 0.05$. Light blue points converge to the segment of fixed points $S^*$, while light red points converge to the unique weird quasiperiodic attractor $\mathcal{A}$.